# Unraveling the Corrosion Mechanism of Boro-Alumino-Phospho-Silicate Glass: Advanced Insights from Solid-State NMR Spectroscopy

Muhammad Amer Khan[a, b], Lili Hu[a], Shubin Chen[a], Yongchun Xu[a], Jinjun Ren[a, b] *

a. Key Laboratory of Materials for High Power Laser, Shanghai Institute of Optics and Fine Mechanics, Chinese Academy of Sciences, Shanghai 201800, P. R. China.

b. Center of Materials Science and Optoelectronics Engineering, University of Chinese Academy of Sciences, Beijing 100049, P. R. China

* Corresponding author.

   E-mail: jinjunren@siom.ac.cn

## Abstract

Corrosion mechanism of minerals and glass is a critical study domain in geology and materials science, vital for comprehending material durability under various environmental conditions. Despite decades of extensive study, a core aspect of these mechanisms—specifically, the formation of amorphous alteration layers upon exposure to aqueous environments—remains controversial. In this study, the corrosion behavior of a boro-alumino-phospho-silicate glass (BAPS) was investigated using advanced solid-state nuclear magnetic resonance (SSNMR) and SEM techniques. The results reveal a uniform nanoscale phase separation into Al-P-rich and Al-Si-rich domains. During corrosion, the Al-P-rich domain undergoes gelation, whereas the Al-Si-rich domain remains vitreous, forming a gel layer comprised of both phases. Although SEM 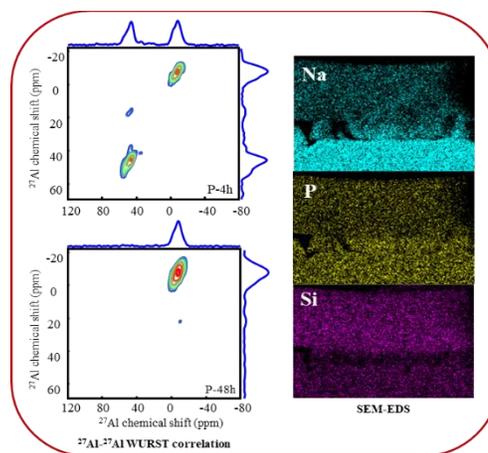 images show a sharp gel/glass interface—suggestive of a dissolution-precipitation mechanism— the phase coexistence within the gel layer provides definitive evidence against such a mechanism. Instead, we propose an in situ transformation mechanism governed by chemical reactions, involving: (i) preferential hydrolysis of Al-P-rich domain leading to porous gel regions; (ii) retention of Al–Si glass domains within the gel layer, with water infiltrating inter-network spaces; and (iii) selective leaching of phosphorus over aluminum, leading to reorganization of the gel network.

# 1. Introduction

Glass has been studied extensively in the science and technological fields [1]. Despite its widespread applications, a significant challenge emerges when glass interacts with aqueous environments, instigating chemical transformations that may undermine its longevity [2]. The interactions between glass and water are influenced by various environmental parameters, resulting in complex alterations over time. This poses a significant challenge, particularly in domains such as nuclear waste management, biomedical instrumentation, and laser optics [3, 4]. The corrosive characteristics of silicate and borosilicate glass have been examined for numerous decades [5, 6]. It is commonly understood that a porous alteration layer, or gel, arises on the surface of the glass when in contact with an aqueous medium, which might have implications for its lasting stability. Several mechanisms have been put forward to address this phenomenon. Among them, the ion-selective diffusion mechanism (IDM) suggests that the alteration layer in alkali silicate glasses results from cation exchange with hydrogen ions when in contact with deionized water [7-9]. Another proposed mechanism, the dissolution-reprecipitation mechanism (DPM), posits that the gel layer is formed by precipitation of various substances on the glass surface [10, 11]. On the other hand, the interfacial dissolution-reprecipitation mechanism (IDPM) suggests that dissolution and gel reprecipitation happen simultaneously at a thin glass interface [12]. Moreover, the multi-step dissolution mechanism (MDM) involves two key processes: (1) ion exchange between mobile cations in glass and H species in the solution, and (2) network dissolution and condensation of soluble elements reaching local saturation, leading to a gel layer on the glass surface [5]. Furthermore, the hydrolysis/condensation mechanism (HCM) suggests that the formation of a gel layer results from the hydrolysis/condensation reaction (or structural rearrangement) that occurs during the glass corrosion process [13]. Despite the significant progress achieved in the aforementioned mechanisms, numerous limitations persist unresolved. IDM attributes the alteration layer in alkali silicate glasses arises from cations exchange and inter-diffusion in deionized water. However, this does not fully elucidate the distinct alteration layer and the notable concentration gradients observed at the interface. Similarly, mechanisms such as DPM, MDM, IDPM, and HCM offer different perspectives on gel layer formation, each presents limitations regarding local saturation and the dynamic processes occurring at the glass-gel interface [14]. However, quantifying dissolution kinetics behind Alteration layer (AL) formation whether AL are formed through the leaching of soluble elements (e.g., Ca, K, Mg, Na) from the primary glass or, on the other hand, through dissolving the primary mineral and precipitating an amorphous Si-rich phase, across a range of varying physicochemical conditions and glass compositions remains a topic of ongoing discussion [15-18].

BAPS glass is pivotal in various advanced technological sectors, including semiconductor manufacturing, sensor development, and nuclear waste management. This glass is utilized particularly in applications such as planarization in large-scale integrated (VLSI) circuit production, solid-state imaging technologies, multilayer chip inductors (MLCI), ion-conducting glass, and medical technologies [19-27]. Understanding the corrosion behavior of this glass in aqueous environments is vital to ensure its long-term stability and efficacy. Previous studies have reported conflicting effects of phosphorus content on glass corrosion behavior. W. M. Paulson et al. [28] observed that increasing phosphorus concentration significantly accelerated corrosion rates, compromising structural integrity. Conversely, Yoshimaru et al. [29] demonstrated that in BAPS glasses containing ≤ 9 wt.% phosphorus, boron facilitates while phosphorus inhibits water penetration, suggesting improved moisture resistance with higher phosphorus content within this compositional range. These contradictory findings highlight the complex, composition-dependent nature of corrosion mechanisms in phosphate-containing glasses.

SSNMR spectroscopy can offer a diverse array of techniques capable of analyzing the structure of glass materials at the atomic scale. This makes it particularly well-suited for studying the outcomes of glass-water reactions. Therefore, in this study, we aim to uncover the corrosion mechanism of glass using SSNMR techniques, with a particular focus on the formation mechanism of the glass gel layer, by analyzing atomic-scale structural changes that occur during the glass corrosion process. We expect the complex structure of this mixed-glass-former glass will provide novel insights into the gel formation process.

## 2. Materials and methods

### 2.1 Glass preparation and leaching experiments

BAPS glass with a nominal composition of $20SiO_2$-$20B_2O_3$-$20P_2O_5$-$7Al_2O_3$-$33Na_2O$ was synthesized using $SiO_2$ (Aladdin, 99.8%), $H_3BO_3$ (Aladdin, 99%), $Al(OH)_3$ (Aladdin, 99.99%), $Na_2CO_3$ (Aladdin, 99.5%), and $Na_3P_3O_9$ (Aladdin, 99.8%) as raw materials. The procedure involved thoroughly mixing 25 g of powder before placing it into a Platinum crucible. The mixture was then melted in a muffle furnace at a temperature of 1200 °C for 40 minutes and quenched at room temperature. The density ($\rho$) of glass is 2.48 g/cm$^3$. Moreover, inductively coupled plasma-atomic emission spectroscopy (ICP-AES) was employed to find the precise elemental quantity of glass as given in Table S1.

Furthermore, the chemical durability of the phosphate glass was assessed using the product consistency test B (PCT-B) [30]. For this, Glass powder was produced via standard methods of grinding, sieving, and ultrasonic cleaning with alcohol, resulting in particles with a diameter range of 75 to 150 μm. For the corrosion experiment, 1 g of glass powder was combined with 10 mL of deionized water in a sealed polytetrafluoroethylene (PTFE) vessel and placed in a furnace at 35°C for durations of 2, 4, 12, 24, and 48 hours. The unaltered glass was labeled as Pristine, while the altered samples were labeled as P-x, (where "P" denotes powdered glass and "x" indicates the exposure duration in hours). Corresponding bulk monolith samples subjected to identical conditions were labeled as B-x, where B represents the bulk sample.

ICP-AES detected the concentration of each leaching element in the leachate. The normalized elemental mass loss ($NL$) can be calculated using the following expression [30]:

$$NL_i = \frac{c_i(sample) - c_i(blank)}{f_i \cdot (SA/V)} = \frac{(c_i(sample) - c_i(blank)) \cdot V \cdot \rho \cdot d}{f_i \cdot 6 \cdot m}$$

Here, $NL_i$ is the normalized mass loss of leached element "$i$". The parameters include $c_i(sample)$ and $c_i(blank)$ are the element "i" concentrations in the glass and glass-free leachate, respectively. The values of $d$ and $V$ were set to 112.5 μm (average powder diameter) and 10 mL (solution volume), respectively. Other parameters include surface area $(SA)$, density of glass $(\rho)$, mass fraction of element "$i$" in the glass $(f_i)$ and glass mass $(m)$.

## 2.2 Solid-State Nuclear Magnetic Resonance (SSNMR) analysis

All SSNMR experiments were performed utilizing a Bruker Avance III HD 500 MHz spectrometer, functioning at a magnetic field strength of 11.7 T and at room temperature. The parameters used in the single pulse experiments and nuclei including $^{27}$Al, $^{31}$P, $^{11}$B, $^{29}$Si, and $^{1}$H are outlined in Table 1. The experimental details and parameters for the 2D NMR analyses, including $^{27}$Al{$^{31}$P} rotational echo double resonance (REDOR), $^{31}$P{$^{27}$Al} rotational echo adiabatic passage double resonance (REAPDOR), $^{27}$Al{$^{1}$H} REDOR, $^{27}$Al{$^{1}$H} 1D $J$-coupling heteronuclear multiple quantum coherence ($J$-HMQC), $^{27}$Al{$^{31}$P} $J$-HMQC, $^{27}$Al{$^{27}$Al} 2D WURST Double Quantum-One Quantum (2Q-1Q), and $^{29}$Si{$^{1}$H} cross polarization (CP) MAS, are provided in the Supporting Information (SI).

Table 1. Experimental parameters used in single pulse MAS NMR analysis.

| Elements | MAS NMR probe (mm) | resonance frequency (MHz) | spinning rate (kHz) | relaxation delay (s) | Pulse Length (μs) | chemical shift standards |
|---|---|---|---|---|---|---|
| $^{31}$P | 4 | 202.45 | 12 | 40 | 3 (90°) | ADP ($\delta$ = 1.12 ppm) |
| $^{27}$Al | 4 | 130.32 | 12 | 0.5 | 0.87 (10°) | 1 M Al(NO$_3$)$_3$ ($\delta$ = 0 ppm) |
| $^{11}$B | 4 | 160.46 | 10 | 8 | 0.8 (15°) | 1M H$_3$BO$_3$ ($\delta$ =19.5 ppm) |
| $^{29}$Si | 4 | 99.35 | 6 | 350 | 1.73 (90°) | TTMS ($\delta$ = −9.7 ppm) |

## 2.3 The experimental details of SEM-EDS

A Hitachi SU8220 field-emission scanning electron microscope was used for microstructural examination. Imaging was carried out in secondary electron (SE) detection mode with an emission current of 10.5 μA, an accelerating voltage of 10 kV, and a working distance of 8.4 mm. The images were obtained at a magnification of 4000× with a 1280 × 960 pixels resolution. Elemental analysis was conducted with an Oxford Instruments X-MA™ 150 mm$^2$ EDS detector under identical working circumstances (10 kV accelerating voltage).

## 3. Results and discussion

### 3.1 Leaching behavior

Figure S1(a) shows the evolution of normalized loss (NL) of glass elements, exhibiting an increasing trend over time. However, the leaching behavior varied across different components in the glass. Boron showed the highest leaching during immersion, followed by sodium and phosphorus. Silicon's normalized mass loss initially increased before stabilizing over time. Furthermore, aluminum slightly declined after an initial rise, as detailed in Table S2. These results highlight the incongruent nature of leaching (as further supported by SEM analysis discussed later).

### 3.2 Static NMR

After immersion, the glass was dried before experimental analysis. The drying effect on glass surfaces can substantially modify their structural and functional properties. Variations in drying conditions, including temperature and environmental exposure, can result in surface modifications that affect performance [31, 32]. However, in the present work, analysis of $^{27}$Al and $^{29}$Si spectra before and after drying, as shown in Figure S1(b, c), reveals no significant variations. Therefore, the drying process did not result in observable structural changes.

### 3.3 MAS NMR analysis

Figure 1(a, b) presents the $^{31}$P NMR spectra of pristine BAPS glass and its corroded counterpart, deconvoluted into four distinct units (I-IV) with chemical shifts at 3.0, -3.1, -8.0, and -14.9 ppm, respectively. The spectra were analyzed using DMFIT software, with fitting parameters detailed in Table S3. Signal III exhibits significantly greater intensity in the corroded sample than in the pristine glass at the cost of signal II. The $P^n_{mAl,xB}$ notation scheme was employed to assign

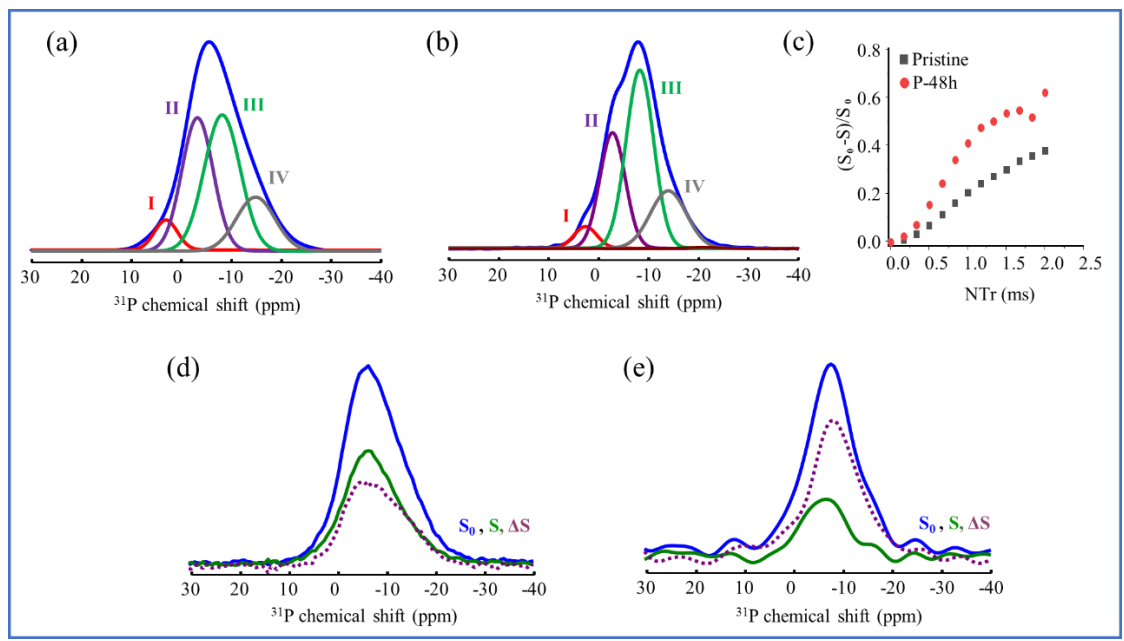

Figure 1(a) single pulse $^{31}$P MAS NMR spectrum of Pristine glass, (b) single pulse $^{31}$P MAS NMR spectrum of corroded glass (P-48h), (c) $^{31}$P {$^{27}$Al} REAPDOR dephasing curve of pristine and P-48h glass. The curves were obtained by integrating the entire spectral region. (d, e) comparison of the $^{31}$P signals with (S) and without (S$_0$) $^{31}$P-$^{27}$Al dipolar interaction from pristine and corroded glass P-48h, acquired using $^{31}$P{$^{27}$Al} REAPDOR experiment at an mixing time of 2 ms, respectively. Whereas, ΔS=S$_0$-S.

the signals, where 'n' denotes the total number of bridging oxygen connections involving P-O-P and P-O-Si bonds, and 'm' and 'x' specify the number of aluminum and boron connections with phosphate, respectively. Based on the chemical shifts, these signals correspond to $P^n_{mAl, xB}$ (n=0 and 1) species. To investigate phosphorus structural evolution during corrosion, we performed $^{31}P\{^{27}Al\}$ REAPDOR experiments (Figure 1c-e). Figure 1c shows the normalized signal attenuation curves at various mixing times for both the pristine and corroded glass. The corroded glass exhibits stronger signal attenuation compared to the pristine glass, indicating that phosphorus tetrahedra are bonded to more $Al^{3+}$ after corrosion. Figures 1d and 1e show the $^{31}P$ spectra acquired with and without recoupling the $^{31}P\leftrightarrow^{27}Al$ dipolar interactions. In addition to the overall stronger attenuation in the corroded glass, the signals on the right (III–IV) show more pronounced attenuation than those on the left (I–II), demonstrating that the phosphorus species associated with signals III and IV are bonded to more $Al^{3+}$ ions than those corresponding to signals I and II. Collectively, these results indicate that after corrosion, phosphorus species are bonded to more $Al^{3+}$. Figure 2(a) presents the $^{11}B$ MAS spectra of pristine and altered (P-2h, P-12h, and P-48h) glass samples. With increasing immersion time, the $B^4$ units of altered glass narrowed significantly,

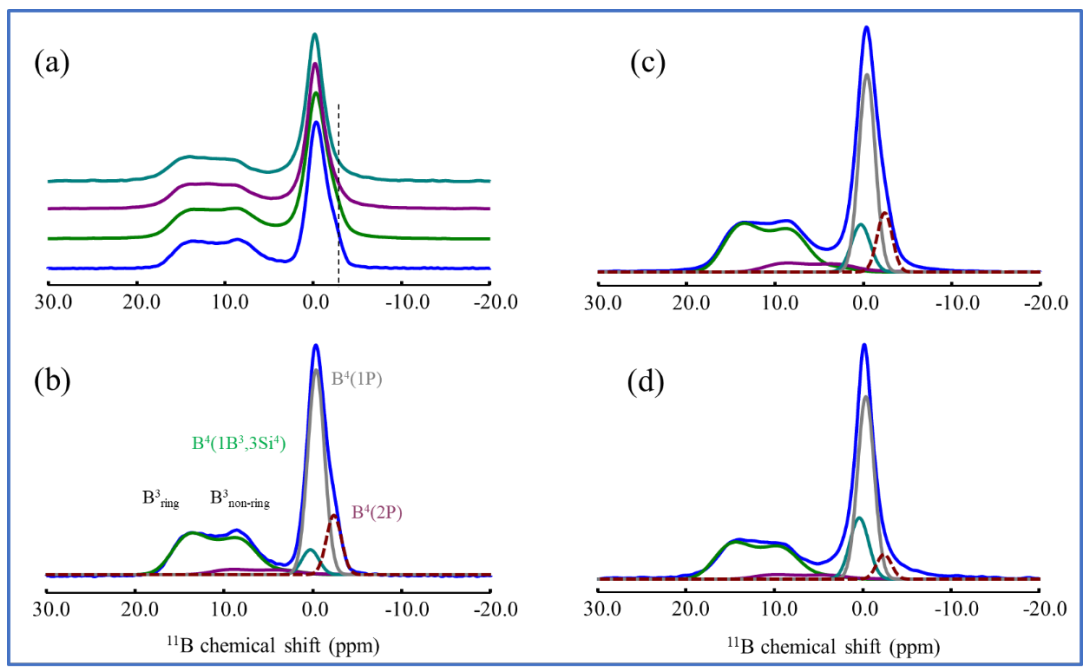

Figure 2(a) compares the $^{11}B$ MAS NMR spectra of the pristine glass and altered glass (P-2h, P-12h, and P-48h) samples, arranged from bottom to top respectively. (b) The $^{11}B$ deconvoluted spectra of pristine glass. (c and d) represent $^{11}B$ spectra of altered glass P-2h and P-48h, respectively.

highlighting the alteration in the local environment. To analyze this, the $^{11}B$ MAS NMR spectrum of pristine glass was deconvoluted using DMFIT, as shown in Figure 2(b), with the deconvolution parameters detailed in Table S4. Based on the literature and chemical shift positions, the broad peaks centered around 17.5 ppm and 13 ppm can be assigned to trigonal ring and non-ring boron units, respectively [33-35]. The chemical shift position of $B^4$ species associated with $^{31}P$ falls within a very narrow range (width < 2 ppm), making it challenging to resolve specific $B^4$(nP) (where nP represents the number of phosphorus linked with boron) units. However, using the shift ranges commonly reported in the literature, the peaks at 0.24 ppm, -0.4 ppm, and -2.4 ppm can be attributed to $B^4(1B^3,3Si^4)$, $B^4(1P)$, and $B^4(2P)$ units, respectively [36-39]. Figure 2(b-d) shows the deconvoluted spectra of pristine, P-2h, and P-48h glass samples, respectively. Notably, the intensity of $B^4(2P)$ unit decreased significantly after the glass was immersed for 48 hours, indicating that interaction with water induced the breakdown of the B-O-P linkage, particularly involving $B^4(2P)$ species, since they have the most B-O-P bonds.

Figure 3(a) illustrates the $^{27}Al$ MAS NMR spectra of pristine and altered glass samples. In pristine glass, a prominent signal, accompanied by two smaller peaks, were detected at chemical shifts of 54.1 ppm, 19 ppm, and -5 ppm, which correspond to aluminum in four-coordinated ($Al^4$), five-coordinated ($Al^5$), and six-coordinated ($Al^6$) configurations, respectively. The $^{27}Al$ chemical shift values were evaluated by analyzing the data through the Czsample model in DMFIT, as shown in Figure S2 (Pristine), and corresponding parameters are given in Table S5. Following immersion, the $Al^4$ peak intensity decreased substantially, while the $Al^6$ signal intensified, reflecting alteration layer formation. Furthermore, the $Al^4$ peak exhibited deshielding, whereas the $Al^6$ chemical shift

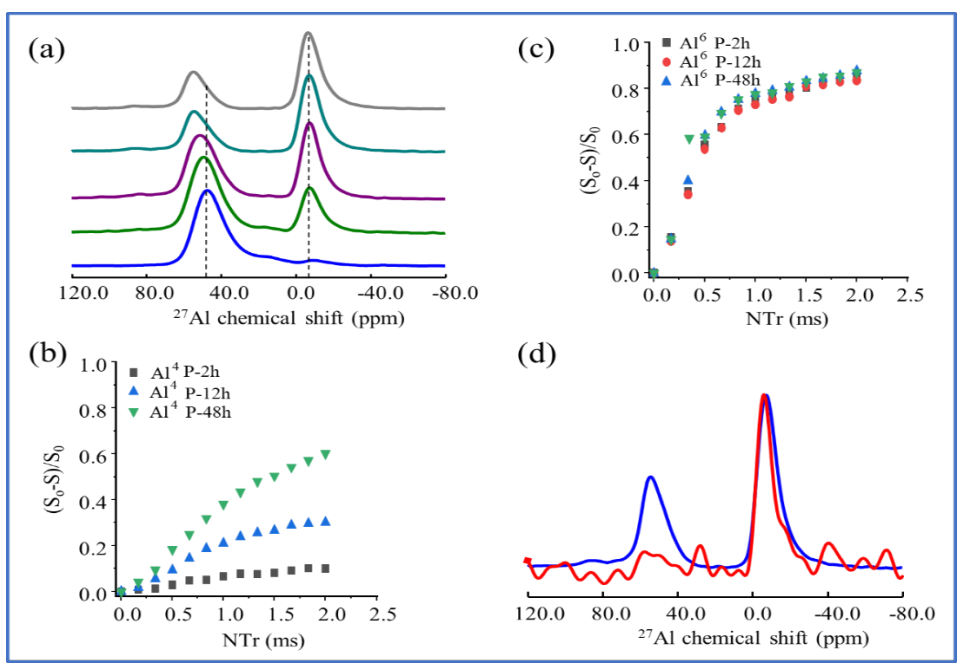

Figure 3(a) presents the $^{27}Al$ MAS NMR spectra of the pristine glass and altered samples (P-2h, P-4h, P-8h, P-12h, and P-48h), displayed from bottom to top. 3(b) and 3(c) depict the $^{27}Al$ {$^1H$} REDOR dephasing curves for the $Al^4$ and $Al^6$ species, respectively. 3(d) shows the $^{27}Al$ {$^1H$} 1D HMQC spectrum (red) with a mixing time of 2.0 ms, alongside the corresponding $^{27}Al$ MAS NMR spectrum (blue) for the P-12h sample.

remained unchanged. This suggests the significant alteration of the $Al^4$ species during corrosion.

To investigate the alteration of Al during the immersion process, the $^{27}Al\{^1H\}$ REDOR and $^{27}Al\{^1H\}$ 1D *J*-HMQC were performed. These experiments provide critical insights into the Al-H interactions in corroded glass, revealing structural variations during the corrosion process. The $^{27}Al\{^1H\}$ REDOR experiment (figure 3b) shows that in early stages of corrosion, $Al^4$ exhibits weak signal attenuation, indicating low hydrogen contents in the surrounding. In contrast, $Al^6$ displays significant REDOR signal attenuation (Figure 3c). As corrosion progresses, the increasing $^{27}Al\{^1H\}$ REDOR signal attenuation for $Al^4$ in samples P-12h and P-48h indicates progressive hydrogen diffusion into the glass. However, the REDOR curve for $Al^6$ remains unchanged over time, suggesting a saturation in hydrogen concentration within this environment, which indicates $Al^6$ exists in the gel layer while $Al^4$ is in the glass phase. The $^{27}Al\{^1H\}$ 1D $J$-HMQC experiment can investigate direct Al-O-H bonding as shown in Figure 3(d). This experiment was performed with two different mixing durations (1.5 ms and 2 ms) to ensure that the result was not affected by the mixing time. The spectrum presented in the figure 3(d) was acquired at 2 ms, whereas the spectrum obtained at 1.5 ms mixing time is not shown here. In both cases, the absence of an $Al^4$ signal in the HMQC spectrum of glass P-12h suggests the lack of direct $Al^4$-O-H connectivity. This indicates water molecules are chemically unreactive with the glass, occupying interstitial positions while preserving network integrity. In contrast, the distinct $Al^6$ peak unequivocally confirms the formation of $Al^6$-O-H bonds, indicating that water molecules cleave the Al-containing network to generate new hydroxylated $Al^6$ species, resulting in a more open and loosened structure. These observations demonstrate that the hydrated glass phase retains a dense, consolidated architecture, whereas the gel layer adopts a highly porous configuration. These structurally distinct difference between the gel layer and glass phase suggests a phase-separated system rather than a continuous gradient, consistent with prior studies on similar glass-water interaction systems [40].

$^{27}Al\{^{31}P\}$ REDOR experiment was employed to investigate the spatial arrangement of phosphorus atoms surrounding aluminum ($Al^4$ and $Al^6$) species, as shown in figure 4(a and b),

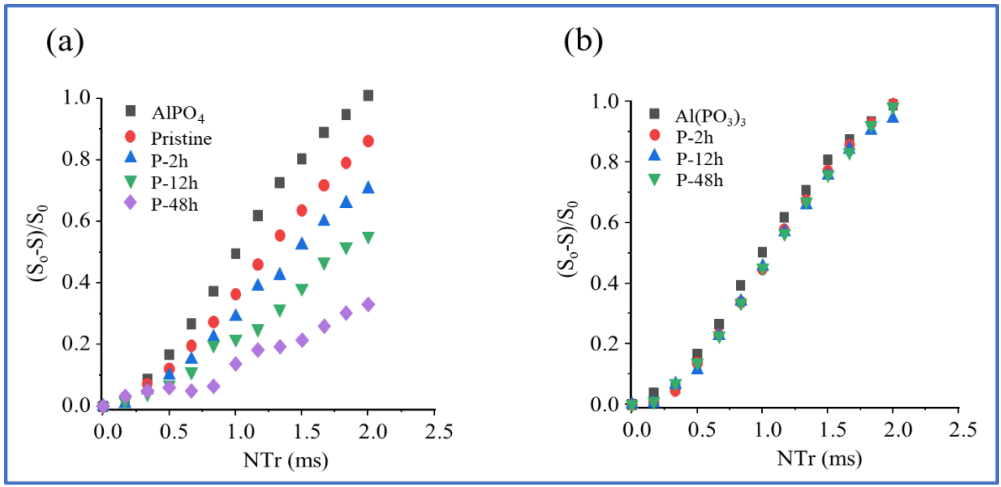

Figure 4(a) and (b) show the $^{27}Al\{^{31}P\}$ REDOR curves for $Al^4$ and $Al^6$, respectively.

respectively. The corresponding $M_2$ (Al-P) for different aluminum species are summarized in Table S5. In pristine glass, the $M_2$ ($Al^4$-P) value is significantly lower than those observed in reference $AlPO_4$ crystals, implying that the $Al^{4+}$ coordination sphere is not exclusively composed of $[PO_4]$ tetrahedra. Upon corrosion, the $M_2$ ($Al^4$-P) values consistently decline. In contrast, the $M_2$ ($Al^6$-P) values remain relatively constant and slightly smaller than that of $Al(PO_3)_3$, implying that the coordination shell of $Al^6$ in the corroded glass layer is almost only [$PO_4$]. The remaining group in the coordination shell of $Al^6$ should be OH, as suggested by the above $^{27}Al\{^1H\}$ 1D *J*-HMQC result.

These results imply the following structure evolution: $Al^4$ has two different species in the pristine glass. One is bonded by silica tetrahedron [$SiO_4$] and the other by [$PO_4$]. When the glass is immersed in water, the $Al^4$ bonded by phosphorus absorbs and reacts with water, changing into $Al^6$, while the $Al^4$ bonded by silica remains unchanged. That is why with the increase of immersion time, $Al^6$ content increases at the cost of $Al^4$ and the $M_2$ ($Al^4$-P) values exhibit a consistent decline. Therefore, it can be concluded that the glass has a high aluminum silicate phase and a high aluminum phosphate phase. During the immersion, the aluminophosphate phase changed into a gel, and silicate glass became a hydrated glass phase. If the aluminosilicate phase were to undergo a hypothetical transformation into a gel, it would adopt a loose and porous structure. The water confined within these pores would be expected to react with $Al^4$ species, generating $Al^6$ and resulting in the presence of substantial $Al^6$ sites connected to silicon. The solid-state NMR signal for such silicon-connected $Al^6$ is well-established, appearing between 0–20 ppm, as extensively documented in aluminosilicate sol-gel systems [41, 42]. However, our $^{27}Al$ NMR spectrum shows a definitive absence of signals in this region. This key observation demonstrates that the aluminosilicate phase does not exist in a gel form but remains a glass. This conclusion is also corroborated by our above $^{27}Al\{^1H\}$ 1D *J*-HMQC and $^{27}Al\{^1H\}$ experiments, which confirm that the $Al^4$ species are embedded within a glassy matrix. Collectively, these results provide consistent evidence that the aluminosilicate phase is a glass, whereas the aluminophosphate phase is a gel.

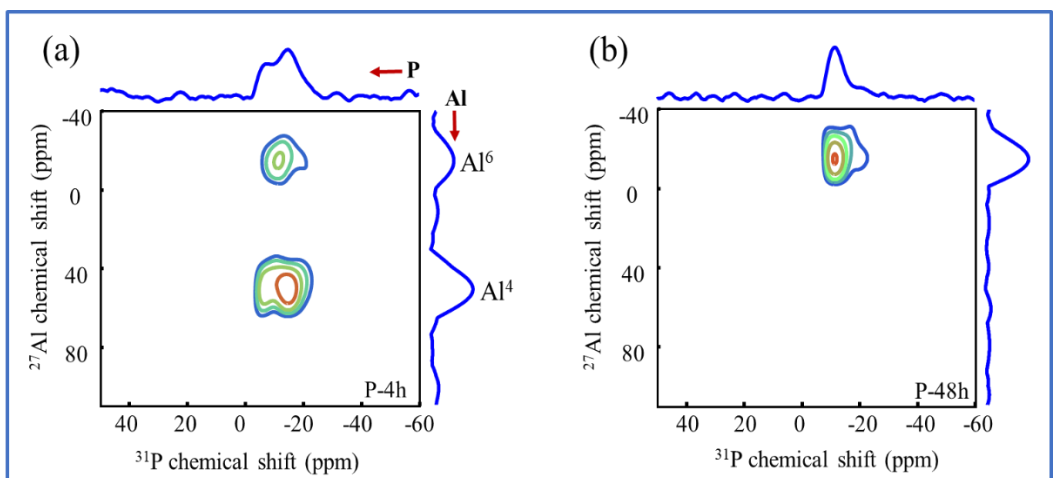

Figure 5(a, b) illustrate the $^{27}Al$ $\{^{31}P\}$ 2D-HMQC experiment for sample P-4h and P-48h, respectively.

$^{27}Al\{^{31}P\}$ 2D-HMQC experiments were done to investigate the glass structure evolution further. The $^{27}Al\{^{31}P\}$ 2D-HMQC spectra present Al-P bond connectivity for the glass P-4h and P-48h, as shown in Figure 5. The $^{27}Al\{^{31}P\}$ 2D-HMQC for P-4h glass (figure 5a) indicates both $Al^6$ and $Al^4$ have P-O-Al bonds connectivity. However, in the sample P-48h (figure 5b), the signal corresponding P-O-$Al^4$ connectivity disappeared, and only P-O-$Al^6$ species were detected on the $^{27}Al\{^{31}P\}$ 2D-HMQC spectrum. These results prove that the $Al^4$ in the phosphate gradually

changes into $Al^6$ during the corrosion, and after 48 hours of immersion, all the $Al^4$ bonding to phosphorus changed into $Al^6$.

The $^{27}Al\{^{27}Al\}$ 2D WURST 2Q-1Q correlation experiments (Figure 6a-c) were conducted to reveal spatial correlations between aluminum species in corroded glasses at sub-nanometer scales. In the P-2h sample, clear $Al^4$-$Al^4$ and $Al^6$-$Al^6$ correlations were observed without any $Al^6$-$Al^4$ cross-peaks. $Al^4$ and $Al^6$ are in the glass phase and gel layer, respectively. No $Al^6$-$Al^4$ cross-peak demonstrates phase segregation between the gel layer and glass phase. With increasing immersion time (P-12h and P-48h), the intensified $Al^6$-$Al^6$ correlations indicated gel phase growth, while diminishing $Al^4$-$Al^4$ correlations reflected consumption of the original aluminum-phosphate clusters (not aluminosilicate phases). The absence of $Al^4$-$Al^4$ correlation in the aluminum silicate phase is likely because the distance between $Al^4$ in the silicate is too far away to create correlation through the $Al^4$-$Al^4$ dipolar-dipolar interaction. This evolution, coupled with disappearing P-O-$Al^4$ connectivity and decreasing $M_2$ ($Al^4$-P) values with the increase of corrosion time, consistently evidences phase separation into phosphate-rich and silicate-rich regions in this glass, with progressive conversion of phosphorus phases into gel during corrosion. The small $M_2$ ($Al^4$-P) value of the P-48h sample, the disappearance of P-O-$Al^4$ connectivity on the $^{27}Al\{^{31}P\}$ 2D-HMQC spectrum, and the vanishing of $Al^4$-$Al^4$ correlations on the $^{27}Al\{^{27}Al\}$ 2D WURST 2Q-1Q spectrum all indicate that in this sample, the aluminum-phosphorus phase within the glass has completely transformed into a gel. In other words, the entire glass has been thoroughly corroded. The remaining $Al^4$ signals originate from the aluminum-silicate glass phase. This means that the material becomes a mixture of phosphate gel and silicate glass phase after glass corrosion.

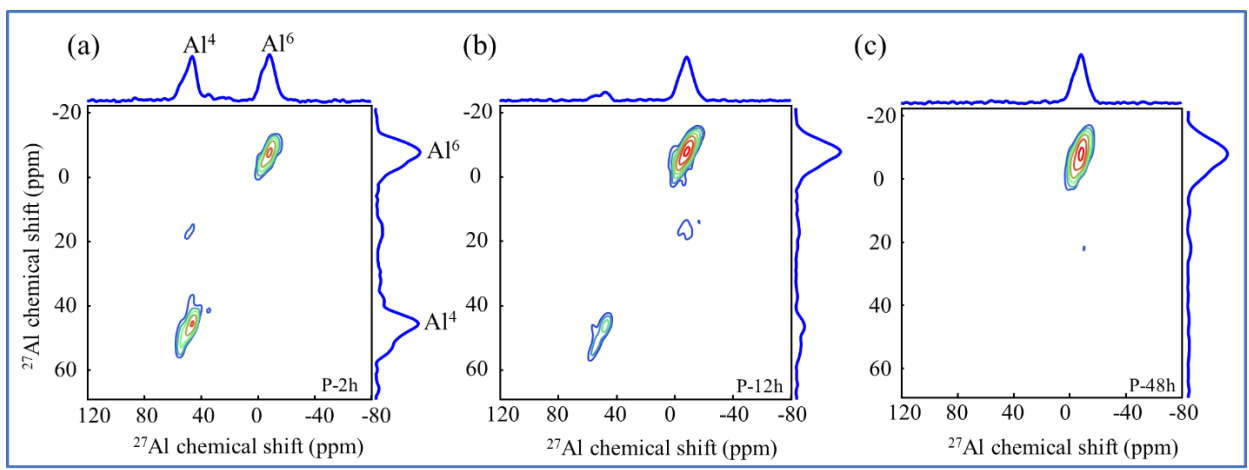

Figure 6(a-c) present the $^{27}Al$-$^{27}Al$ 2D WURST 2Q-1Q correlation spectra of the corroded glass P-2h, P-12h and P-48h, respectively.

The $^{27}Al\{^{31}P\}$ 2D-HMQC results prove that all the aluminum phosphorus phase has changed into gel, and the $Al^4$ in the sample P-48h is only the Al left in the silicate phase. This can facilitate the precise localization and bandwidth of $Al^4$ signal from the silicate phase in the spectra and help to deconvolve the $^{27}Al$ NMR spectra for both pristine and corroded glass, as illustrated in Figure S2, while fitting parameters are given in Table S5. The $Al^4$ signals were separated into two distinct peaks, consisting of $Al^4$ bonding to phosphorus and silicate. As corrosion time increases, peak-II ($Al^4$ bonding to phosphorus) progressively diminishes, indicating the transformation of $Al^4$ in the glass to $Al^6$ in the gel layer. The peak-I is associated with Al-O-Si linkage, which remains stable for a longer duration. This behavior reinforces a model of homogeneous phase separation in BAPS glass at the nano-scale, wherein aluminum engages in P-O-Al and Si-O-Al linkages across separate P-rich and Si-rich domains. The preferential transformation of the phosphorus-rich phase into gel arises from its lower aqueous stability. Structurally, the aluminum phosphorus phase primarily comprises $P^0$ and $P^1$ species, whereas $Q^2$ and $Q^3$ dominate the silicate phase (see below). The higher density of the silicate network impedes water penetration, while the more open phosphorus-rich phase facilitates hydrolytic reactions and gel formation.

**$^{29}$Si NMR analysis**

Figure 7(a) compares the $^{29}$Si spectra of pristine and altered glass. The signal-to-noise ratio (SNR) is impaired by the low silicon content and long relaxation time, which inherently restricts overlapping in the $^{29}$Si MAS spectra, making deconvolution even more challenging. The spectral weight center is approximately at -98 ppm; this chemical shift falls within the range between the typical chemical shifts of $Q^2$ and $Q^3$ species, indicating that the silicate species are predominantly $Q^2$ and $Q^3$. Such broad $^{29}$Si spectra suggest the presence of more than two signals, potentially including $Q^1$ and $Q^4$. However, deconvolution of these species is extremely difficult due to severe overlapping and poor SNR. There is no significant difference in the $^{29}$Si spectral weight center

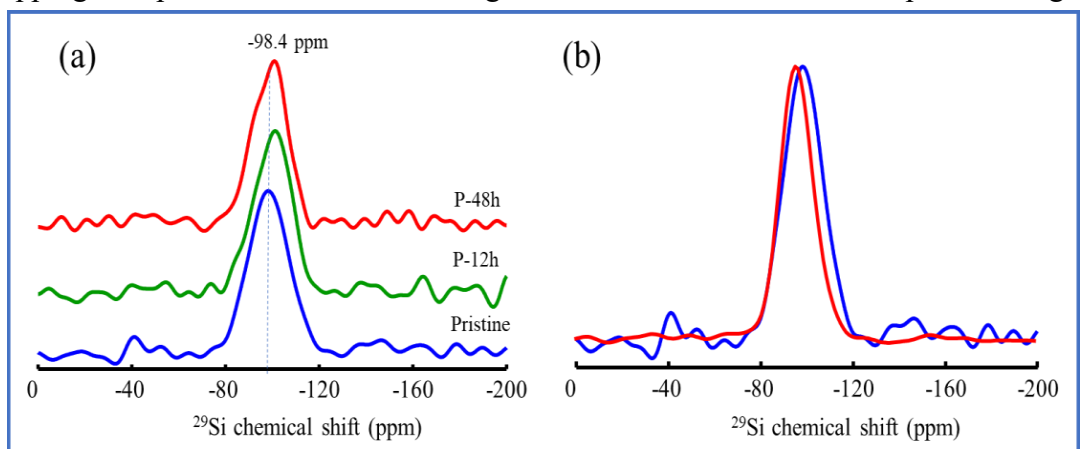

Figure 7(a) shows the $^{29}$Si single pulse MAS spectra of pristine and corroded glass, (b) compares CP MAS spectra (red) and $^{29}$Si single pulse (blue) MAS spectra of P-48h glass.

between the pristine and corroded glass (figure 7a), implying that the silicate glass network remains largely unbroken. In contrast, the $^{29}$Si{$^1$H} CP MAS spectrum (figure 7b) exhibits a distinct shift toward higher chemical shifts, which indicates that hydrogen species have incorporated into the silicate phase and are enriched in loosely structured silicate species such as $Q^2$ and $Q^1$.

**Bulk glass analysis**

Since the powder sample does not readily allow observation of the gel layer or the interface between the gel and the glass phase, we prepared bulk glass to directly examine the gel layer on the surface. To gain a comprehensive understanding of the elemental distribution across the gel layer and the glass phase, we conducted similar corrosion experiments on the bulk glass and performed Scanning Electron Microscopy (SEM) coupled with Energy Dispersive X-ray Spectroscopy (EDS) analysis on the corroded monolith, as shown in Figure 8. For this, bulk glass possessed dimensions of (~1.66 cm × 1.54 cm × 0.15 cm), immersed in 60 ml of solution for 2 hours (named as B-2h) under identical experimental conditions as those applied to glass powder. The SEM-EDS mapping reveals distinct differences in the elemental distribution between the gel and pristine regions, providing insights into the leaching behavior of key elements. In the gel layer, sodium exhibited significant depletion, indicated by the lighter coloration, which confirms its high mobility and rapid leaching from the glass surface. Phosphorus, on the other hand, also showed

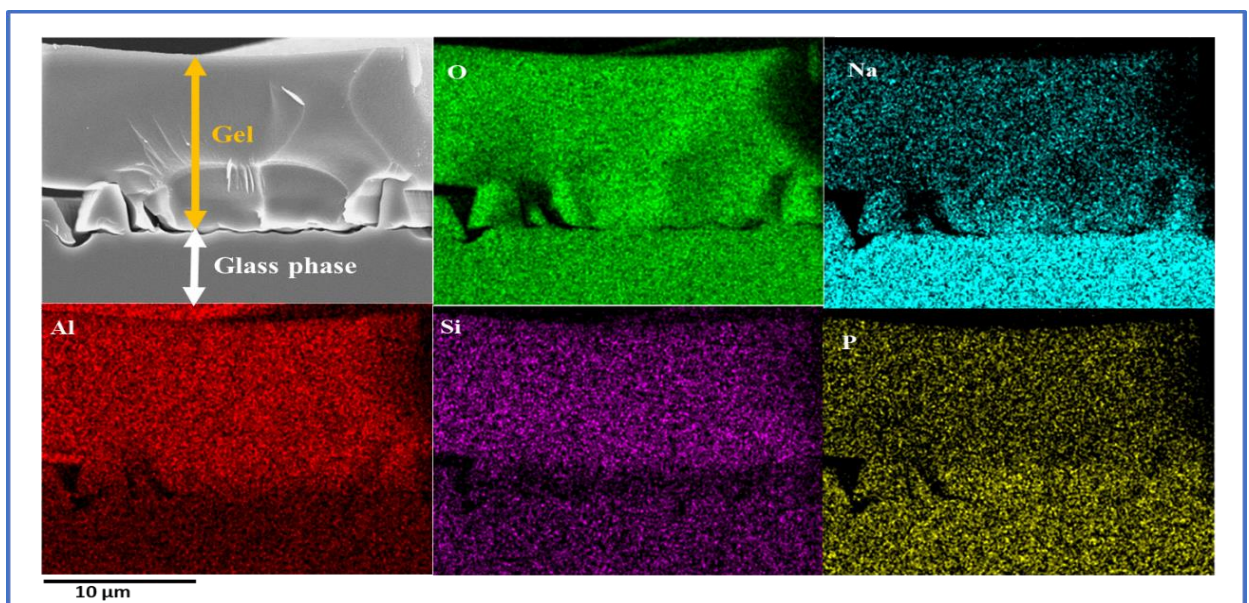

Figure 8 presents the SEM-EDS results of B-2h glass sample.

noticeable depletion. SEM image clearly shows that a distinct gel layer approximately 13 µm thick is separated from the pristine glass phase, with a sharp compositional transition at the interface that is consistent with findings reported in the literature [11, 43, 44]. Collectively, these results align with the conclusions derived from $^{27}$Al-$^{27}$Al 2D-WURST correlation analyses of glass powder samples, which confirm that the gel layer is phase-separated from the glass phase. For extended corrosion period 48 hours (B-48h), optical microscopy tracked the growth of an increasingly fragile gel layer whose mechanical instability prevented SEM characterization. As illustrated in Figure 9(a), a notable gel layer is visible on the glass surface after 48 hours of corrosion, providing an overall view of the altered glass. Moreover, Figure 9(b) shows the gel layer thickness for glass B-48h, measuring 156 µm, which indicates ongoing corrosion over time. To investigate the structure of the gel layer on the bulk glass, the layer was peeled off the glass substrate for NMR measurements. Although the gel layer can be readily detached from the bulk glass due to cracking at the gel-matrix interface (Figures 8 and 9a), trace amounts of glass material remain adhered to the peeled gel layer. However, in extensively corroded samples like B-120h, the bulk glass completely transforms into gel without residual glass fragments. Figure 10(a) presents a comparative analysis of $^{27}$Al MAS NMR spectra between the corroded powder glass (P-48h) and the gel layer derived from bulk glass (B-48h), revealing similar spectral features. Consistent with the 1D spectra, the $^{27}$Al-$^{27}$Al WURST 2D correlation (Figure 10b and c) analysis reveals parallel evolution in the powder glass and the gel layer of the bulk glass systems: initial coexistence of Al$^4$-Al$^4$ and Al$^6$-Al$^6$ correlations (B-48h; Figure 10b) followed by exclusive Al$^6$-Al$^6$ correlations after prolonged immersion (B-120h; Figure 10c). Likewise, the 1D $J$-HMQC spectra (Figure 10d) confirm the absence of Al$^4$-O-H connectivity and the exclusive presence of Al$^6$-O-H bonds. For

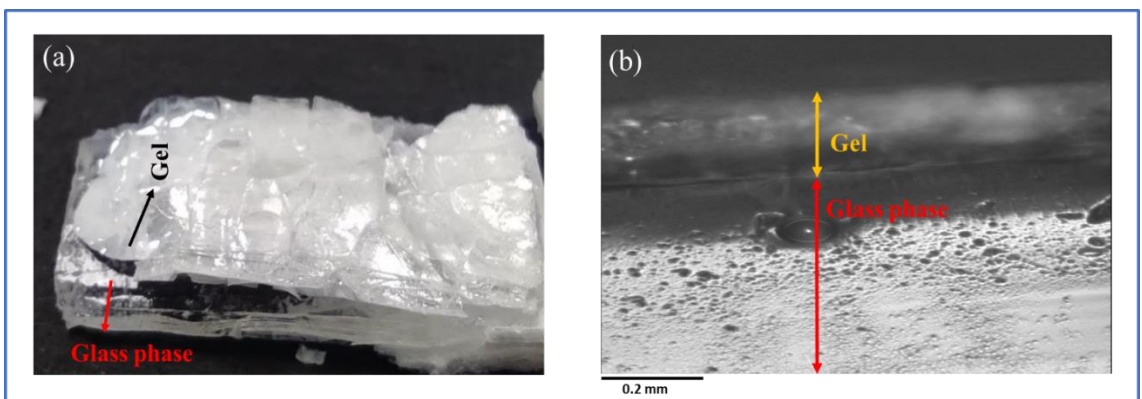

Figure 9(a) provides a broad view of corroded bulk glass B-48h while (b) shows the gel layer thickness at 10X magnification for glass B-48h.

glass samples (B-48h and B-120h), the $^{27}$Al{$^1$H} REDOR analysis (Figure S3-a, b) indicates the gradual hydration of Al$^4$ and a saturated hydrogen environment for Al$^6$. Furthermore, the lack of Al$^4$-O-P connectivity in $^{27}$Al{$^{31}$P} 2D-HMQC (Figure S3-d) indicates the complete transformation of phosphorus phase into gel. These consistent observations reveal that the gel layer of the bulk glass comprises coexisting glass and gel phases, mirroring the behavior observed in the corroded powder glass. This finding is critical, as it demonstrates the dual-phase nature of the gel layer (glass + gel). Notably, the persistence of a glass phase within the gel layer challenges conventional dissolution-reprecipitation mechanisms (e.g., DPM/IDPM), which would predict a purely gel-like product. Instead, these results strongly support an alternative transformation pathway dominated by in situ network reorganization rather than complete dissolution-reprecipitation. The results demonstrate that the aluminum-phosphorus gel phase exhibits a more open structure than the glass phase. This structural porosity promotes preferential leaching of P and Na, resulting in their significant depletion within the gel phase relative to the glass matrix. Notably, as the aluminum-phosphorus gel constitutes the dominant phase in the gel layer, its selective gelation - even while the aluminum silicate remains glass phase - is sufficient to establish a sharp compositional gradient across the gel-matrix interface at submicron scale. The $^{31}$P MAS NMR spectrum of sample B-48h exhibits an upfield chemical shift after corrosion (Figure 10e), suggesting increased P-Al bonding. This observation aligns with the $^{31}$P NMR data from powdered glass (P-48h). This can be attributed to the higher P leaching rate than Al, which results in an elevated Al/P ratio in the corroded glass. The chemical shift further implies densification through enhanced Al-P interactions during corrosion.

Figure 10(f) presents the $^{29}$Si MAS NMR spectra for pristine and corroded B-48h glass. A significant positive shift following corrosion suggests silicate network partial depolymerization during the corrosion. However, the powder glass (P-48h) shows almost no change in the $^{29}$Si spectrum. This contrasting behavior of silicon between the powder and bulk samples should be due to the differences in SA/V. To investigate the effect of solution saturation on the silicate network, we analyzed three glass powder samples with systematically varied surface-area-to-volume (SA/V) ratios after 48-hour corrosion: P-48h (SA/V = 2108 m$^{-1}$), P-48h(a) (175 m$^{-1}$), and P-48h(b) (38 m$^{-1}$). The SA/V ratios were adjusted by increasing the solution volume while keeping the glass surface area constant. As anticipated, lower SA/V conditions (e.g., 38 m$^{-1}$) delay solution saturation due to greater dilution of reaction products, thereby prolonging the corrosion phase. ICP data (Table S6) confirm this mechanism, showing a ~9-fold increase in silicon leaching at SA/V =

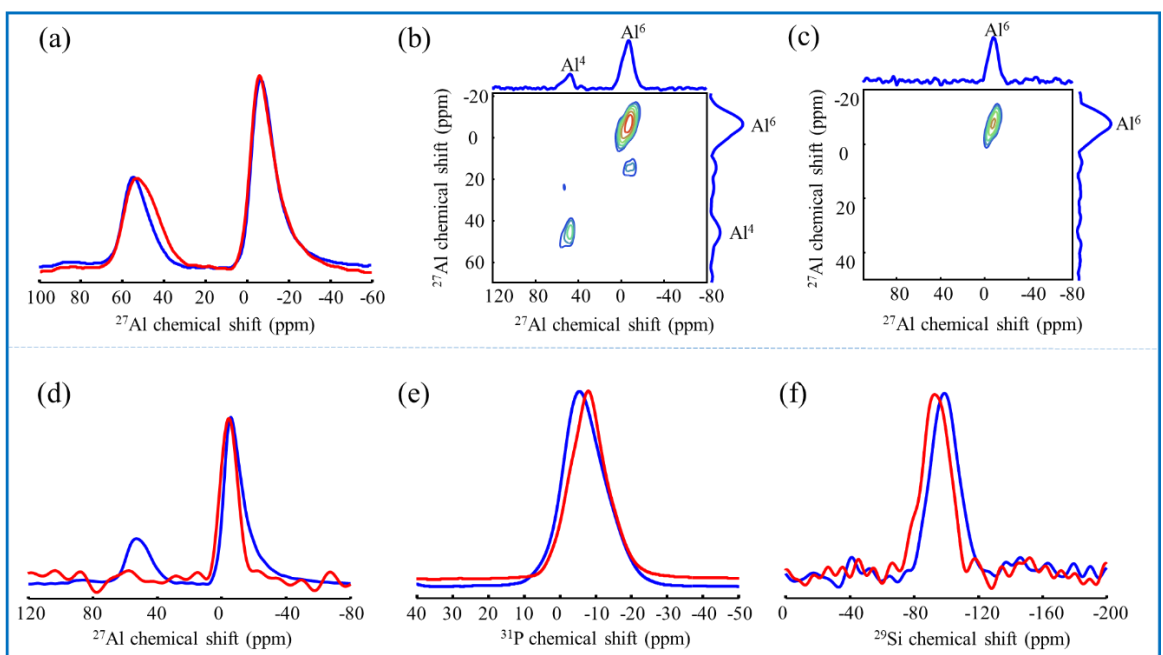

Figure 10(a) presents the single-pulse $^{27}$Al MAS NMR spectra of samples P-48h (blue) and B-48h (red). Figures 10(b) and 10(c) show the $^{27}$Al-$^{27}$Al WURST correlation spectra for glasses B-48h and B-120h, respectively. Figure 10(d) displays the single-pulse $^{27}$Al MAS spectrum of glass B-48h (blue) alongside the $^{27}$Al-$^{1}$H 1D HMQC spectrum (red). Figure 10(e) compares the $^{31}$P NMR spectra of the pristine glass (blue) and B-48h (red), while Figure 10(f) presents the $^{29}$Si NMR spectra of the pristine (blue) and B-48h (red) glass samples.

38 m$^{-1}$ compared to high SA/V (2108 m$^{-1}$) over 48 hours. The $^{29}$Si NMR spectra (Figure S4) further support this trend. For sample P-48h(b) (SA/V = 38 m$^{-1}$), a positive chemical shift is observed, suggesting partial depolymerization of the silicate network. In contrast, sample P-48h (SA/V = 2108 m$^{-1}$) shows only a negligible shift, indicating minimal dissolution of silica. This spectral divergence aligns with the ICP results: At low SA/V, the dominance of downfield peaks reflects extended corrosion, while high SA/V conditions induce rapid saturation, decreasing the dissolution and suppressing spectral changes. Critically, while the SA/V ratio decreases substantially, the leaching rates of all elements show enhancements, which is clear evidence that the solution stays far from saturation under low SA/V regimes. The continued formation of gel layers under these non-saturation conditions (low SA/V) provides definitive evidence against precipitation-based formation mechanisms in another way.

Although above results indicate that while mutual diffusion of water and glass components occurs during corrosion, the formation of the gel layer is not diffusion-controlled; that is, the corrosion rate is not limited by the diffusion process. The key mechanism lies in water molecules preferentially attacking hydrolysis-sensitive bonds in the glass network, causing localized structural damage that creates a more open network capable of accommodating more water. As local water concentration increases, progressive network degradation occurs, enabling the gel layer to advance inward. Meanwhile, water diffusing deeper into the glass remains at insufficient concentrations to disrupt the network, preserving the dense glass structure and preventing rapid component diffusion/leaching. This process creates a sharp compositional interface between the gel layer and pristine glass. Furthermore, chemical reactions within the gel layer induce structural reorganization, leading us to define this as a chemical-reaction-controlled in situ transformation mechanism.

## 4. Conclusion

In this study, we investigated the corrosion mechanism of a boro-alumino-phospho-silicate glass system. Although we observed a gel layer on the glass surface with a distinct, abrupt compositional transition between the gel layer and the pristine glass matrix using SEM- a feature seemingly consistent with dissolution-precipitation models, our advanced solid-state NMR results conclusively demonstrate that the gel layer consists of coexisting glass and gel phases, providing definitive evidence against dissolution-reprecipitation as the formation process. Furthermore, while the overall corrosion involves mutual diffusion of water and glass components, the gel layer formation is not diffusion-controlled. We define this novel gel layer formation mechanism as a chemical-reaction-controlled in situ transformation mechanism.

The mechanism underlying the formation of the gel layer on the glass surface is as follows: Water molecules first preferentially attack hydrolysis-sensitive bonds in the glass network (in this glass, it is B-O-Si, B-O-P, P-O-Al, and P-O-P bonds) on the surface, causing localized structural damage that creates a more open network capable of accommodating more water. As local water concentration increases, progressive network degradation occurs preferentially along the aluminum-phosphate phase, ultimately converting the entire aluminum-phosphate phase into gel. The gel exhibits selective leaching behavior, with $P^{5+}$ being preferentially removed over $Al^{3+}$. The residual $Al^{3+}$ subsequently reacts with $[PO_4]^{3-}$ groups, partially reconstructing the damaged gel network. In contrast, the aluminum-silicate phase remains a glass phase. The resulting gel layer, therefore, represents an intermixed composite of the transformed aluminum-phosphate gel phase and the intact aluminum-silicate glass phase.

### Data availability

Data is provided within the manuscript or supplementary information files. Raw data can be provided upon reasonable request.

## Acknowledgment

The author (Muhammad Amer Khan) gratefully acknowledge the financial support provided by the CAS-ANSO Scholarship.

## Author contributions

Jinjun Ren conceived and designed the research project, and interpreted the data. Muhammad Amer Khan prepared the glass and conducted the measurements except for the ICP-AES and SEM-EDS experiments, and interpreted the data. Yongchun Xu conducted the ICP-AES experiment. Jinjun Ren and Muhammad Amer Khan wrote the manuscript with inputs from Lili Hu, Yongchun Xu and Shubin Chen. All the authors discussed the results.

## Funding Declaration

This work was supported by the Strategic Priority Research Program of the Chinese Academy of Sciences [Grant No. XDB0650000].

## Competing interests

The authors declare that they have no competing interests.

# Supplementary information

# Unraveling the Corrosion Mechanism of Boro-Alumino-Phospho-Silicate Glass: Advanced Insights from Solid-State NMR Spectroscopy

Muhammad Amer Khan[a, b], Lili Hu[a], Shubin Chen[a], Yongchun Xu[a], Jinjun Ren[a, b] *

a. Advanced Laser and Optoelectronic Functional Materials Department, Special Glasses and Fibers Research Center, Shanghai Institute of Optics and Fine Mechanics, Chinese Academy of Sciences, Shanghai 201800, P. R. China.

b. Center of Materials Science and Optoelectronics Engineering, University of Chinese Academy of Sciences, Beijing 100049, P. R. China

* Corresponding author.

  E-mail: jinjunren@siom.ac.cn

## Experimental details

**i.** REDOR/REAPDOR experiments: Atomic spatial distribution and dipolar interaction strength were studied using rotational echo Double resonance (REDOR) studies. Applying π pulses on the influenced nuclei allows REDOR to restore dipolar interactions, averaged by MAS. A normalized REDOR ($\Delta S/S_0$) REDOR curve was obtained as a function of dipole evolution time (NTr). All experiments used a 4 mm MAS probe at 12 kHz, with a saturation comb added before relaxation. The typical pulse sequence of Gullion and Schaefer, as modified by Chan and Eckert, was used in the $^{27}$Al{$^{31}$P} REDOR experiments [1, 2]. For both the $^{27}$Al and $^{31}$P nuclei, the 180° pulse lengths were 8.5 μs. Moreover, the relaxation time was 1s. A short evolution time ($0 \leq \Delta S/S_0 \leq 0.2$) was used to obtain the second dipolar moment, or $M_2^{SI}$ ($S=^{27}$Al, $I=^{31}$P), from the REDOR parabolic curve by using the following equation:

$$\frac{\Delta S}{S_0} = \frac{1}{I(I+1)\pi^2}(NT_r)^2 M_2^{SI}$$

$\Delta S/S_0 = (S_0 - S)/S_0$, where $S_0$ and $S$ are the signal intensities measured with and without dipolar recoupling effects. This ratio was calculated as a function of dipolar evolution time, denoted as NTr, where N indicates the number of rotor cycles and Tr denotes the length of the evolution period. The $^{31}$P{$^{27}$Al} REAPDOR experiments were carried out using a pulse sequence developed by Garbow and Gullion [3]. In $^{31}$P{$^{27}$Al} REAPDOR pulse sequence, for the $^{31}$P nuclei, a pulse length of 8.5 μs (180°) and a relaxation delay of 40 s were set. For $^{27}$Al, the recoupling pulse was 27.8 μs, corresponding to one-third of the rotor period, and the nutation frequency was set to 22.1 kHz, determined using a liquid sample of Al(NO$_3$)$_3$. Furthermore, the $^{27}$Al{$^1$H} REDOR experiments employed the Gullion and Schaefer pulse sequence [3]. The relaxation time for $^{27}$Al{$^{1}$H} REDOR was 1.0 s, and 180° pulse duration was 7.0 μs.

**ii.** $^{27}$Al{$^{1}$H} 1D *J*-HMQC and $^{27}$Al{$^{31}$P} *J*-HMQC: The $^{27}$Al{$^{1}$H} 1D heteronuclear double quantum coherence (HMQC) experiment was performed utilizing a 4 mm magic angle spinning (MAS) probe operating at a rotation frequency of 12 kHz and a relaxation interval of 1.0 s, accompanied by saturation comb before the experiment. To detect heteronuclear J-coupling, the heteronuclear double quantum coherence was excited by two 90° pulses across the $^{27}$Al and $^{1}$H nuclei, each with a pulse length of 4 μs. This method directly detects bonding connectivity between Al and H nuclei, providing direct evidence of their chemical bonds [4]. WURST-enhanced $^{31}$P{$^{27}$Al} *J*-HMQC tests were conducted to probe direct P-O-Al connectivity, utilizing a 4 mm MAS probe rotating at 12 kHz. The WURST-80 excitation pulse was utilized on the Al channel to enhance the sensitivity of the $^{31}$P{$^{27}$Al} *J*-HMQC. The scan rate of the WURST-80 excitation pulse was synchronized with the spinning speed of 12 kHz. The pulse duration was fine-tuned to 250 μs, while the deviation from the central frequency was fine-tuned to ±350 kHz. The mixing period, denoted as $\tau$, for generating *J*-coupled double quantum coherence was optimized to 4.75 μs, during which the magnetization underwent evolution. The duration of the 90° pulses for $^{31}$P and $^{27}$Al nuclei was 4.25 μs and 10 μs, respectively. To ensure stable signal acquisition, a saturation comb pulse was added to the P channel before a relaxation period of 60 s.

**iii.** $^{27}$Al{$^{27}$Al} 2D WURST 2Q-1Q (Double Quantum-One Quantum) correlation: The $^{27}$Al{$^{27}$Al} 2D WURST 2Q-1Q correlation experiment was conducted to investigate the proximity between Al species [5]. In this experiment, the BR221 pulse technique developed by Wang et al. [6], was employed with WURST to increase signal intensity. Furthermore, this experiment was performed using a 4 mm MAS probe with a spinning rate of 12 kHz. Excitation and reconversion periods were set to 1666.7 μs. The $^{27}$Al pulse lengths at 90° and 180° were 14.5 μs and 29.0 μs, respectively. A saturation comb was used before a 1.0 s relaxation interval. Following the double transformation, the correlated signals in the F1 dimension are positioned half the total of their chemical shifts in the F2 dimension.

**iv.** $^{29}$Si{$^{1}$H} CP MAS: This experiment examined the interaction between Si and H atoms. The experiment used the TPPM15 pulse scheme with a 160° pulse length of 4.44 μs on the H channel for decoupling during data acquisition. A relaxation delay of 1.0 s was applied and the nutation frequencies of $^{1}$H and $^{29}$Si were 90.9 and 32.2 kHz, respectively. The contact time was 3.0 ms, and the H power level underwent a linear ramp down from 90.9 kHz to a nutation frequency of 45.5 kHz.

# Supplement Results

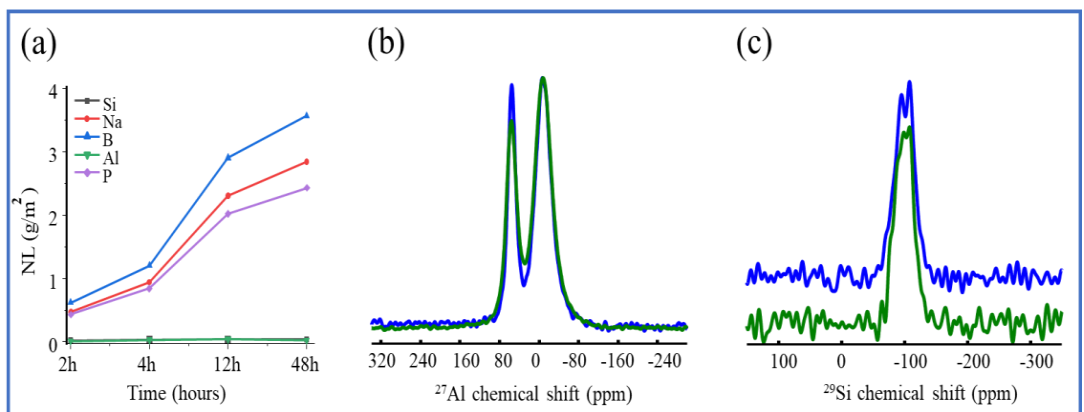

Figure S1(a) shows normalized leaching at different immersion times. (b) Static $^{27}$Al NMR spectra of wet (green) and dry (blue) P-48h sample. (c) Static $^{29}$Si spectra of wet (green) and dry (blue) P-48h sample.

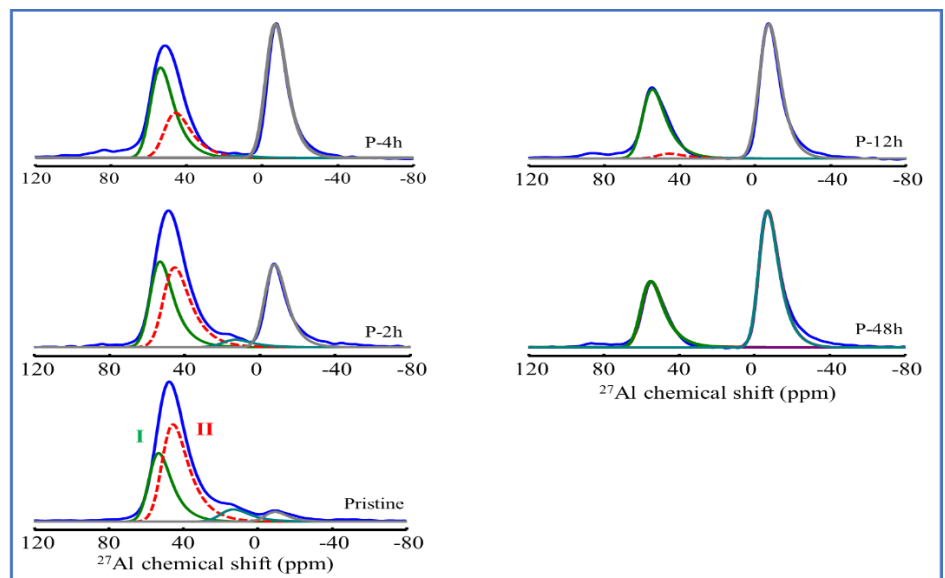

Figure S2 presents the deconvoluted $^{27}$Al NMR spectra of pristine and corroded glass samples.

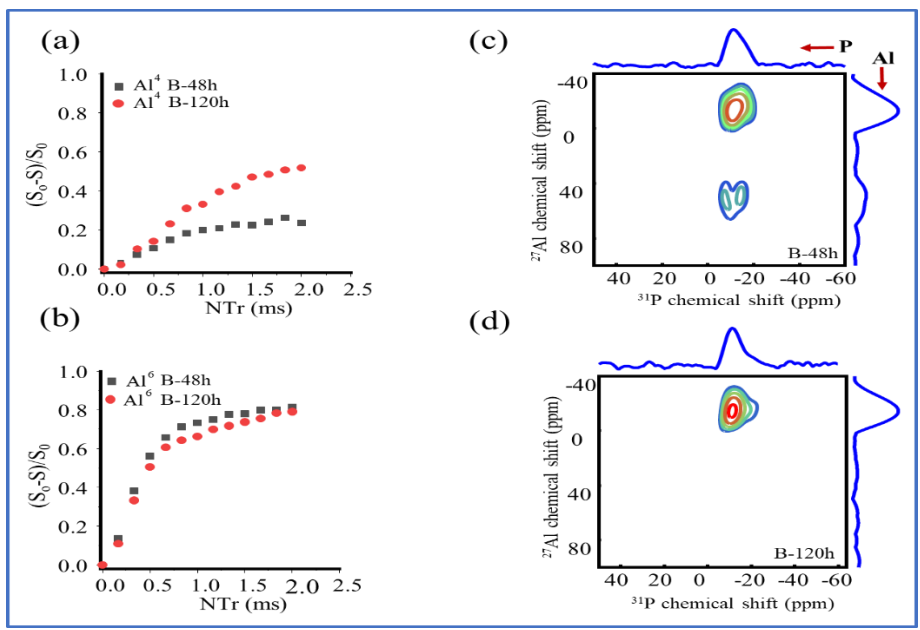

Figure S3(a, b) show the $^{27}$Al{$^{1}$H} REDOR curves for the Al$^4$ and Al$^6$ units, while (c, d) present the corresponding $^{27}$Al{$^{31}$P} 2D-HMQC spectra of samples B-48h and B-120h.

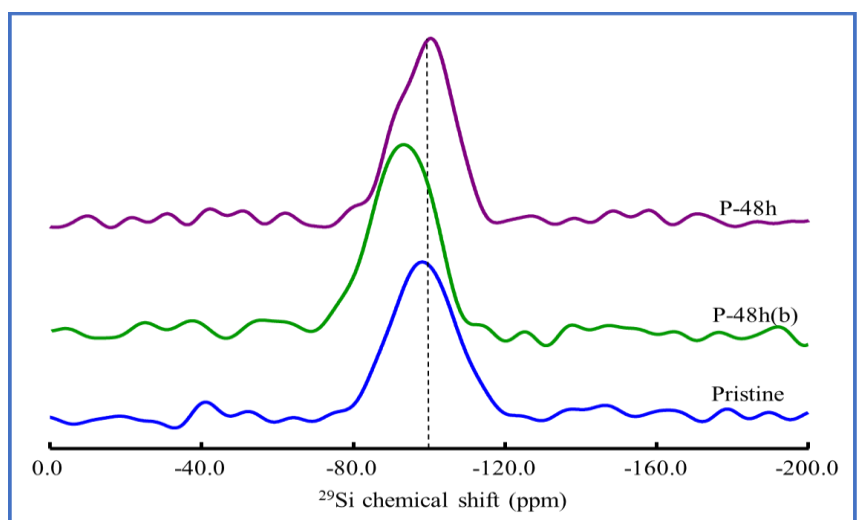

Figure S4 compares the $^{29}$Si NMR spectra of pristine, P-48h(b) at SA/V=38 m$^{-1}$ and P-48h at SA/V=2018 m$^{-1}$ of glass powder samples.

Table S1 presents nominal wt. % and with corresponding ICP-AES measurements in parenthesis.

| Glass | $Al_2O_3$ wt. % (±2%) | $P_2O_5$ wt. % (±2%) | $Na_2O$ wt. % (±2%) |
|---|---|---|---|
| BAPS (pristine) | 4.61(4.72) | 15.12(14.49) | 18.51(18.95) |

Table S2 normalized mass loss for elements present in BAPS glass.

| Time (hours) | NL (g/m$^{-2}$) | | | | |
|---|---|---|---|---|---|
| | B (± 0.9) | Na (± 1.2) | P (± 0.7) | Al (± 0.1) | Si (± 0.1) |
| 2 h | 6.1 | 4.6 | 4.4 | 0.1 | 0.2 |
| 4 h | 12.0 | 9.1 | 8.8 | 0.3 | 0.3 |
| 12 h | 29.0 | 22.6 | 21.1 | 0.3 | 0.4 |
| 48 h | 35.6 | 27.8 | 25.3 | 0.2 | 0.4 |

Note: Errors in parenthesis represent the maximum standard deviation from duplicate experiments.

Table S3 includes the $^{31}$P fitting parameters of pristine glass and altered glass (P-48h, values in parenthesis).

| Peaks | $\delta_{iso}^{cs}$ [ppm(±0.3)] | FWHM [ppm (±0.3)] | Area [% (±3)] |
|---|---|---|---|
| I | 3.0 (3.0) | 5.4 (5.4) | 5.3 (3.1) |
| II | -3.1 (-2.7) | 7.4 (7.0) | 31.5 (27.9) |
| III | -8.0 (-8.5) | 8.5 (7.7) | 42.3 (53.5) |
| IV | -14.9 (-14.9) | 9.1 (9.1) | 21.2 (15.5) |

Table S4 includes the fitting parameters of $^{11}$B units for pristine and corroded BAPS glass samples.

| Name | unit | $\delta_{iso}^{cs}$ [ppm (±0.3)] | Area [% (±2)] |
|---|---|---|---|
| Pristine | $B^3 ring$ | 17.5 | 36.3 |
| | $B^3\ non\ ring$ | 13.0 | 5.0 |
| | $B^4(1B^3, 3Si^4)$ | 0.25 | 5.3 |
| | $B^4(1P)$ | -0.4 | 42.2 |
| | $B^4(2P)$ | -2.4 | 11.1 |
| P-2h | $B^3 ring$ | 17.5 | 37.0 |
| | $B^3\ non\ ring$ | 13.4 | 7.5 |
| | $B^4(1B^3, 3Si^4)$ | 0.3 | 9.8 |
| | $B^4(1P)$ | -0.4 | 36.0 |
| | $B^4(2P)$ | -2.4 | 10.0 |
| P-48h | $B^3 ring$ | 18.5 | 39.92 |
| | $B^3\ non\ ring$ | 14 | 4.71 |
| | $B^4(1B^3, 3Si^4)$ | 0.4 | 15.4 |
| | $B^4(1P)$ | -0.35 | 40.0 |
| | $B^4(2P)$ | -2.4 | 4.9 |

Table S5 shows the isotropic chemical shift ($\delta_{iso}$), quadrupolar coupling constant ($C_Q$), the fraction of $^{27}$Al species (%), and the dipolar second moment $M_2$ (Al-P) obtained from the $^{27}$Al {$^{31}$P} REDOR spectra for both glass (pristine + corroded) and crystalline glass reference.

| Samples | Units | $\delta_{iso}$ [ppm(±0.3)] | $C_Q$ (MHz) | Fraction [%(±2)] | $M_2^{Al-P\ a}$ ×10$^6$rad$^2$/s$^2$ | $M_2^{Al-P\ b}$ ×10$^6$rad$^2$/s$^2$ |
|---|---|---|---|---|---|---|
| Pristine | Al(IV)$^I$ | 58.0 | 4.3 | 33.3 | 3.4 | - |
| | Al(IV)$^{II}$ | 51.0 | 4.7 | 54.7 | | |
| | Al(V) | 19.0 | 4.7 | 7.5 | | |
| | Al(VI) | -5.0 | 3.9 | 4.5 | | |
| P-2h | Al(IV)$^I$ | 58.0 | 4.3 | 32.5 | 2.7 | 3.8 |
| | Al(IV)$^{II}$ | 51.0 | 4.7 | 35.0 | | |
| | Al(V) | 19.0 | 4.7 | 3.42 | | |
| | Al(VI) | -3.4 | 3.9 | 29.0 | | |
|  | Al(IV)$^I$ | 58.5 | 4.3 | 34.2 |  |  |
|  | Al(IV)$^{II}$ | 51.3 | 4.7 | 19.9 | nm | nm |
| P-4h | Al(V) | 19.0 | 4.7 | 1.0 |  |  |
|  | Al(VI) | -3.2 | 3.7 | 44.8 |  |  |
|  | Al(IV)$^I$ | 59.3 | 4.3 | 35.5 |  |  |
|  | Al(IV)$^{II}$ | 51.2 | 4.7 | 3.0 |  |  |
| P-12h | Al(V) | 19.0 | 4.7 | 0.5 | 2 | 3.5 |
|  | Al(VI) | -3.2 | 3.7 | 61.0 |  |  |
|  | Al(IV)$^I$ | 59.6 | 4.3 | 35.5 |  |  |
|  | Al(IV)$^{II}$ | - | - | - |  |  |
| P-48h | Al(V) | 19.0 | 4.7 | 0.4 | 1.2 | 4.1 |
|  | Al(VI) | -2.9 | 3.9 | 64.1 |  |  |
|  |  |  |  |  | AlPO$_4$(5.0) | Al(PO$_3$)$_3$(4.7) |

'a' and 'b' denote $M_2$ for Al$^4$-O-P and Al$^6$-O-P units, respectively. While 'nm' denotes not measured.

Table S6 presents the normalized mass loss for the BAPS glass corroded for 48 hours at different SA/V values.

| SA/V (m$^{-1}$) | Normalize loss NL(g/m$^2$) | | | | |
|---|---|---|---|---|---|
|  | B (± 0.8) | Na (± 0.9) | P (± 1.0) | Al (± 0.1) | Si (± 0.1) |
| 2108 | 35.6 | 28.5 | 24.3 | 0.3 | 0.4 |
| 175 | 38.4 | 31.6 | 27.9 | 0.2 | 1.5 |
| 38 | 38.0 | 31.4 | 28.9 | 0.2 | 3.7 |